\documentclass[conference]{IEEEtran}
\IEEEoverridecommandlockouts
\usepackage{cite}
\usepackage{amsmath,amssymb,amsfonts}
\usepackage{algorithmic}
\usepackage{graphicx}
\usepackage{textcomp}
\usepackage{xcolor}

\usepackage{amsthm}
\usepackage{float}
\usepackage{tikz}
\usepackage{comment}
\usepackage{mathrsfs}

\newtheorem{setn}{Theorem}[section]

\newtheorem{cor}[setn]{Corollary}

\theoremstyle{definition}

\newtheorem{exa}[setn]{Example}

\renewcommand{\mod}[1]{\ (\textnormal{mod }{#1})}
\newcommand{\suco}[2]{\left.#1\right|_{\mathbb{F}_{#2}}}
\newcommand{\Fqm}{\ensuremath{\mathbb{F}_{q^m}}}
\newcommand{\Fq}{{\ensuremath{\mathbb{F}_{q}}}}
\newcommand{\Sch}{{\ensuremath{\mathscr{S }\,}}}
\DeclareMathOperator{\tr}{Tr}

\begin{document}

\title{Low-Complexity PIR Using Subfield Subcodes\\

}

\author{\IEEEauthorblockN{Christian J. Lex
}
\IEEEauthorblockA{\textit{Department of Mathematical Sciences,} \\
\textit{Aalborg University}\\
Aalborg, Denmark \\
clex16@student.aau.dk
}
\and
\IEEEauthorblockN{Oliver W.~Gnilke
}
\IEEEauthorblockA{\textit{Department of Mathematical Sciences,} \\
\textit{Aalborg University}\\
Aalborg, Denmark \\
owg@math.aau.dk}
}

\maketitle

\begin{abstract}
A major drawback of many PIR schemes is the high computational cost at the servers. We present a scheme that uses only operations in the prime field during response generation. For binary extension fields this leads to schemes that only need XOR operations at the servers to calculate the responses. This is achieved by restricting the queries to a subfield subcode or trace code. 
We investigate possible parameter ranges and focus on the example of GRS codes and subfield subcodes of these.

\end{abstract}


\section{Introduction}
\IEEEPARstart{A}{} lot of work on PIR considers only two parameters of interest, the download cost, or rate, and storage overhead. The effort at the server side, which can be significant is usually ignored. Two exceptions are \cite{Julien}, where a PIR scheme that requires no computation on the server side is presented. And \cite{Eitan}, which introduces a new parameter called access complexity, that measures how many files a server has to access to calculate its response. 

In this paper we will present the use of subfield subcodes to a private information retrieval (PIR) scheme. In particular, we consider an alteration of the PIR scheme presented in \cite{Gnilke} which uses generalised Reed-Solomon (GRS) codes to achieve a PIR rate of $(n-(k+t-1))/n$ for $n$ servers using a dimension $k$ storage code. This scheme offers protection against $0<t\leq n-k$ colluding servers. By choosing the retrieval code as a subfield subcode of a GRS code (such subfield subcodes are at times called alternant codes) instead of a GRS code a significant computational cost is saved during file retrieval. We compare the complexity of this subfield subcode scheme to the scheme in \cite{Gnilke} using both GRS codes over any field of characteristic $2$ as well as GRS codes over $\mathbb{F}_{q^m}$ where $q>2$.

 \section{Private Information Retrieval}
We use the setup in \cite{Gnilke}. Let $n$ denote the number of servers in the DSS, let $\mu$ denote the number of files, and let $b$ denote the number of rows in each file. The files $x^1,\hdots,x^\mu\in (\mathbb{F}_{q^m})^{b\times k}$ are encoded row-wise using a linear $[n,k,d_C]_{q^m}$ code $C$, the \textit{storage code}, as
\begin{align}
    Y = \begin{bmatrix}x^1\\\vdots\\x^\mu\end{bmatrix}G_C,
\end{align}
 where $G_C$ is a generator matrix for $C$. Superscripts will refer to files, subscripts to servers, and parenthesis to vector entries. E.g., $y^i = x^iG_C$ denotes the $i^{\textnormal{th}}$ encoded file, $\boldsymbol{y}^{i}_j$ the part of this encoded file on the $j^{\textnormal{th}}$ server, and $y^i_j(a)$ the $a^{\textnormal{th}}$ entry of this vector.
 
Now, the PIR scheme is described. We will denote this scheme by $\mathscr{S}$. Let $D$ be a linear length $n$ code over $\mathbb{F}_{q^m}$. We call this code the \textit{retrieval code}. The star product of the storage code and the retrieval code is defined as 
\begin{align}\label{eq:star}
    C \star D = \textnormal{span}\left\{(c_0d_0,\hdots, c_{n-1}d_{n-1}) \ \middle| \ \boldsymbol{c}\in C,\boldsymbol{d} \in D\right\}
\end{align}
 Let $c := d_{C\star D}-1$. The set $J := \{1,\hdots,\max\{k,c\}\}$ will be the index set of servers (after perhaps a rearrangement of servers) from which symbols are downloaded. To make sure that exactly one file is downloaded after $s$ iterations of the algorithm we enforce that $b = \textnormal{lcm}(c,k)/k$ and $s = \textnormal{lcm}(c,k)/c$.
 
A query is constructed by choosing $\mu b$ codewords $\boldsymbol{d}^{l,a}$ of $D$ uniformly at random where $l\in [\mu]$ and $a\in [b]$. For each $j\in[n]$ we define a vector $\boldsymbol{d}_j$ by 
\begin{align}
    \boldsymbol{d}_j^l=\left(d^{l,1}(j),\hdots,d^{l,b}(j)\right),\  \boldsymbol{d}_j = \left(\boldsymbol{d}_j^{1},\hdots,\boldsymbol{d}_j^\mu\right)\in (\mathbb{F}_{q^m})^{\mu b}.
\end{align}
A subset $J_1 = [d_{C\star D} -1]\subseteq J$ is defined and is partitioned according to the rows of the files:
\begin{align}\label{eq:partition}
    J_1^1=\left[c/b\right], J_1^2 = \left[2c/b\right]\backslash [c/b],\hdots, J_1^b = [c]\backslash [(b-1)c/b].
\end{align}
Suppose that we wish to fetch the $i^{\textnormal{th}}$ file. Then the $j^{\textnormal{th}}$ query $\boldsymbol{q}_j^i$ is defined by
\begin{align}\label{eq:query}
    \boldsymbol{q}_j^i = \begin{cases}\boldsymbol{d}_j + \boldsymbol{e}_{b(i-1) + a} &\textnormal{if} j \in J_1^a,\\ \boldsymbol{d}_j &\textnormal{if} j \notin J_1.\end{cases}
\end{align}
The $j^{\textnormal{th}}$ entry of the response vector is then determined as 
\begin{align}\label{eq:response}
    r_j^i = \langle\boldsymbol{q}^i_j,\boldsymbol{y}_j\rangle.
\end{align}
This is iterated by applying a length $c/b$ cyclic shift to the partition \eqref{eq:partition} within $J$. That is, if the $a^{\textnormal{th}}$ index set for the $u-1^{\textnormal{th}}$ iteration is $J_{u-1}^a = \{j_1,\hdots,j_{c/b}\}$ then 
\begin{align}
    J_u^a = \left\{j_1 + c/b \mod{|J|},\hdots,j_{c/b} + c/b \mod{|J|}\right\}.
\end{align}
Then the queries are defined as in \eqref{eq:query} now using the partition $J_u^a$ and $J_u$ as the union of the $J_u^a$. The data can now be reconstructed as follows: Choose a parity check matrix $H$ for $C\star D$. Suppose that $\boldsymbol{r}^i$ is the response vector of the first iteration. Then
\begin{align}\label{eq:matvecprod}
    H\boldsymbol{r}^i = H\boldsymbol{c} + H\begin{bmatrix}y^i_1(1)\\\vdots\\y_c^i(b)\\0_{(n-c)\times 1}, \end{bmatrix}
\end{align}
where $c\in C\star D$. Hence, $y_1^i(1),\hdots,y^i_c(b)$ can be recovered from \eqref{eq:matvecprod} since any choice of $c$ columns of $H$ is a linearly independent set. This is done similarly for the other $s-1$ iterations until an entire file can be reconstructed.

 We present here two theorems of \cite{Gnilke}:
\begin{setn}\label{setn:rate}
Let $C$ be an $[n,k,d_C]$-code and let $D$ be some length $n$ retrieval code. If the minimum distance of $C\star D$ denoted by $d_{C\star D}$ satisfies $d_{C\star D}\leq k$ or there exists $J\subseteq [n]$ such that every size $k$ subset of $J$ is an information set of $C$ then $\mathscr{S}$ retrieves the correct file with rate $(d_{C\star D}-1)/n$.
\end{setn}
\begin{setn}\label{setn:protection}
$\mathscr{S}$ protects against $d_{D^{\perp}}-1$ colluding servers.
\end{setn}
 
\section{Low-Complexity Variant}
We will now modify the scheme \Sch to reduce the complexity of calculating the responses $r_j^i$.

The main idea is to replace the retrieval code $D$ with its subfield subcode $D|_{\Fq}$. We have the immediate inclusion $C \star D \supseteq C \star D|_\Fq$ hence we can retrieve at least as many symbols as before. The protection against collusion relies on the min. distance of a trace code.

\begin{setn}[Delsarte \cite{Delsarte}]\label{setn:delsarte}
For a code $D$ over \Fqm, \[ (D|_\Fq)^\perp= \tr(D^\perp). \]
\end{setn}

In general min. distances of trace codes are not easy to determine and even good lower bounds are not known. But they are easy enough to determine for the examples of interest in the present application.

We collect these observations in the following theorem.

\begin{setn}
Let $C$ and $D$ be a storage and a retrieval code respectively for the scheme \Sch. Then we can define a low complexity variant $\Sch|_\Fq$ by replacing $D$ with $D|_\Fq$.
The scheme $\Sch|_\Fq$ has at least the same rate as the scheme \Sch and protects against $t'$-collusion, where $t'=d_{\tr(D^\perp)}-1$.
\end{setn}

We will keep our focus on the variant of the scheme $\mathscr{S}$ using GRS codes as storage and retrieval codes as well as its corresponding subfield subcode scheme $\suco{\mathscr{S}}{q}$. Hence, by $\textnormal{GRS}_k(\boldsymbol{\alpha},\boldsymbol{v})$ we denote the GRS code
\begin{align}
\begin{split}&\textnormal{GRS}_k(\boldsymbol{\alpha},\boldsymbol{v}) = \\
    &\left\{\left(v_0f(\alpha_0),\hdots,v_{n-1}f(\alpha_{n-1})\right)\ \middle| \ f\in \mathbb{F}_{q^m}[X],\deg f < k  \right\},
    \end{split}
\end{align}
for $\boldsymbol{v}\in (\mathbb{F}_{q^m}^*)^n$ and \textit{support} $\boldsymbol{\alpha}\in (\mathbb{F}_{q^m})^n$ satisfying $\alpha_i\neq \alpha_j$ for $i\neq j$.
The GRS codes are closed under taking duals. In particular,
\begin{align}\label{eq:dualgrs}
(\textnormal{GRS}_k(\boldsymbol{\alpha},\boldsymbol{v}))^\perp = \textnormal{GRS}_{n-k}(\boldsymbol{\alpha},\Tilde{\boldsymbol{v}}),
\end{align}
where 
\begin{align}\label{eq:dualtwist}
    \Tilde{\boldsymbol{v}} = \left(\frac{1}{v_0\prod_{i\neq 0}(\alpha_0-\alpha_i)},\hdots, \frac{1}{v_{n-1}\prod_{i\neq n-1}(\alpha_{n-1}-\alpha_i)}\right).
\end{align}
The star product as defined in \eqref{eq:star} for GRS codes $C$ and $D$ satisfies
\begin{align}
    \begin{split}
        \textnormal{GRS}_{k}(\boldsymbol{\alpha}&,\boldsymbol{u}) \star \textnormal{GRS}_{l}(\boldsymbol{\alpha},\boldsymbol{v})=\\
        &\textnormal{GRS}_{\textnormal{min}\{n,k+l-1\}}(\boldsymbol{\alpha},(u_0v_0,\hdots,u_{n-1}v_{n-1})).
    \end{split}
\end{align}
Assume now that we have $D = \textnormal{GRS}_t(\boldsymbol{\alpha},\boldsymbol{v})$ for some $\boldsymbol{\alpha}\in (\mathbb{F}_{q^m})^n$ and $\boldsymbol{v}\in (\mathbb{F}_{q^m}^*)^n$.  Theorem \ref{setn:delsarte}, \eqref{eq:dualgrs}, and \eqref{eq:dualtwist} combines to yield the diagram in Fig. \ref{fig:sfscgrs}.
    \begin{figure}[H]
     \centering
     \begin{tikzpicture}[scale=0.83]
     \node at(-3,1) (1) {$D = \textnormal{GRS}_{t}(\boldsymbol{\alpha},\boldsymbol{v})$};
     \node at (3,1) (2) {$\textnormal{GRS}_{n-t}(\boldsymbol{\alpha},\tilde{\boldsymbol{v}}) = D^{\perp}$};
     \node at (-3,-1) (3) {$\suco{D}{q}=\textnormal{GRS}_t(\boldsymbol{\alpha},\boldsymbol{v})$};
     \node at (3,-1) (4) {$\textnormal{Tr}(\textnormal{GRS}_{n-t}(\boldsymbol{\alpha},\tilde{\boldsymbol{v}})) = \textnormal{Tr}(D^\perp)$};
     \node at (-0.2,1.2) (5) {Dual};
     \node at (-0.4,-0.8) {Dual};
     \node at (-2.5,0) {$\cap \mathbb{F}_q^n$};
     \node at (2.6,0) {Tr};
     \draw[<->,thick] (1) -- (2);
     \draw[->,thick] (1) -- (3);
     \draw[->,thick] (2) -- (4);
     \draw[<->,thick] (3) -- (4);
    \end{tikzpicture}
     \caption{Relation between duals of GRS codes and their subfield subcodes.}
     \label{fig:sfscgrs}
 \end{figure}
Taking the storage code as $C = \textnormal{GRS}_{k}(\boldsymbol{\alpha},\boldsymbol{u})$ and the retrieval code as $\suco{D}{q}$ yields a PIR rate of at least 
\begin{align}
    \frac{n-k-t+1}{n}
\end{align}
and a protection against $d_{\textnormal{Tr}(D^\perp)}-1$ colluding servers assuming that $t\leq n-k.$

In case a lower bound on the collusion protection is required we can assign $\textnormal{Tr}(D) = \textnormal{Tr}(\textnormal{GRS}_{t}(\boldsymbol{\alpha},\boldsymbol{v}))$ as the retrieval code. This yields a collusion protection $t' = d_{\suco{D^\perp}{q}}-1 \geq t$, however in general we have
\begin{align}
    C\star \textnormal{Tr}(D)\not\subseteq C\star D.
\end{align}
Hence, this approach will yield no immediate lower bound on the rate of the new scheme. We collect these observations in the following corollary
\begin{cor}\label{tracescheme}
Let $C$ and $D$ be a storage and a retrieval code respectively for the scheme \Sch. Then we can define a low complexity variant $\tr(\Sch)$ by replacing $D$ with $\tr(D)$.
The scheme $\tr(\Sch)$ has rate $(d_{C \star \tr(D)}-1)/n$ and protects against $t'$-collusion, where $t'=d_{\suco{D^\perp}{q}}-1$.
\end{cor}

\section{Complexity}\label{sec:Complexity}
This scheme has the evident disadvantage to the GRS scheme in \cite{Gnilke} that we have no immediate lower bound on the collusion protection. However, a considerable reduction in time complexity can be achieved by choosing the retrieval code as a subfield subcode. We consider the case $q^m=2^m$. In the scheme $\mathscr{S}$ the response $\boldsymbol{r}^i$ has $n$ entries, hence, in each iteration $n$ inner products \eqref{eq:response} are calculated. Each of these consists of $\mu b-1$ additions and $\mu b$ multiplications in $\mathbb{F}_{2^m}$. An addition in $\mathbb{F}_{2^m}$ has time complexity $\mathcal{O}(m)$ and a multiplication in $\mathbb{F}_{2^m}$ has time complexity $\mathcal{O}(m^2)$ using naïve multiplication. Therefore, determining $\boldsymbol{r}^i$ has complexity 
\begin{align}
    n(\mu b -1)\mathcal{O}(m)+n\mu b\mathcal{O}(m^{2}) = n\mu b\mathcal{O}(m^{2}),
\end{align}
which is iterated $s$ times.

Let us turn to the case where $\suco{D}{2}$ is the retrieval code. Then $\boldsymbol{q}_j^i \in (\mathbb{F}_2)^{\mu b}$, thus, we have only additions in \eqref{eq:response} and determining $\boldsymbol{r}^i$ has complexity
\begin{align}
    n(\mu b-1)\mathcal{O}(m).
\end{align}

We again consider the scheme $\mathscr{S}$ now for $q>2$. Determining $\boldsymbol{r}^i$ consists of $n(\mu b-1)$ additions in $\mathbb{F}_{q^m}$ each of complexity $\mathcal{O}(m\log_2(q))$ and $n\mu b$ multiplications each of complexity $\mathcal{O}(m^2(\log_2(q))^2)$. Hence, each iteration of the response computation has complexity
\begin{align}
 n\mu b\mathcal{O}(m^2(\log_2(q))^2).
\end{align}
In the subfield subcode case the inner product \eqref{eq:response} yields $\mu b -1 $ additions in $\mathbb{F}_{q^m}$ each with complexity $\mathcal{O}(m\log_2(q))$ and $\mu b$ multiplications of an element of $\mathbb{F}_{q^m}$ and an element of $\mathbb{F}_q$ each with complexity $\mathcal{O}(m\log_2(q)^2)$. Hence, we get time complexity for computing $\boldsymbol{r}^i$ as
\begin{align}
   n\mu b \mathcal{O}(m(\log_2(q)^2).
\end{align}
Faster multiplication algorithms than naïve multiplication exist for large integers (see for example \cite{vsGathen}) however the naïve case suffices for our analysis since the fields of consideration are relatively small. 
\section{Examples}
As a proof of concept as well as for clarification we will look at a few small examples. 
\begin{exa}
For the storage code $C_1$ we choose the $[4,1,4]$ repetition code $C_1 = \textnormal{GRS}_1(\boldsymbol{1},\boldsymbol{\alpha}_1)$ for $\boldsymbol{\alpha}_1$ as all of $\mathbb{F}_4$ in any order. The retrieval code is chosen as the Reed-Solomon code $D_1 = \textnormal{GRS}_3(\boldsymbol{1},\boldsymbol{\alpha}_1)$ which has the $[4,1,4]$ repetition code as its dual. Hence, $\suco{D_1}{2}$ has parity check matrix $H = \left[1 \ 1 \ 1 \ 1\right]$, and $\suco{D_1}{2}$ has minimum distance $2$ and dimension $3$ just as $D_1$. This yields a scheme $\suco{\mathscr{S}_1}{2}$ with rate $R_1 = 1/4$ and collusion protection $3$. Thus, in this case, we can pass to the subfield subcode scheme for free.
\end{exa}

As a slightly more interesting example where proper storage coding can be done is as follows.

\begin{exa}
Choose for $k\leq 3$ the storage code as the length $8$ code $C_2 = \textnormal{GRS}_k(\boldsymbol{v}_2,\boldsymbol{\alpha}_2)$ where $\boldsymbol{v}_2$ is any element of $(\mathbb{F}_8^*)^8$ and $\boldsymbol{\alpha}_2$ is all of $\mathbb{F}_8$. The retrieval code is $D_2 = \textnormal{GRS}_5(\boldsymbol{1},\boldsymbol{\alpha}_2)$, and $\suco{D_2}{2}$ has generator matrix
\begin{align}
    \suco{G_2}{2} = \begin{bmatrix}1 & 0 & 0 & 0 & 1 & 1 & 0 & 1\\
                                 0 & 1 & 0 & 0 & 1 & 0 & 1 & 1\\
                                 0 & 0 & 1 & 0 & 0 & 1 & 1 & 1\\
                                 0 & 0 & 0 & 1 & 1 & 1 & 1 & 0 \end{bmatrix}.
\end{align}
We see that $\suco{D_2}{2}$ is the self-dual $[8,4,4]$ extended binary Hamming code. Hence, the scheme $\suco{\mathscr{S}_2}{2}$ protects against $t'=3$ collusion, since $\left(\suco{D_2}{2}\right)^{\perp} = \suco{D_2}{2}$, compared to $t=5$ for the scheme $\mathscr{S}$. Computations show that the rate for both schemes is given by $R_2 = (4-k)/8$.
\end{exa}
At last, let us consider a trinary example. 
\begin{exa}
Thus, we take the storage code $C_3 = \textnormal{GRS}_k(\boldsymbol{v}_3,\boldsymbol{\alpha}_3)$ for some $\boldsymbol{v}_3\in (\mathbb{F}_9^*)^9$, $\boldsymbol{\alpha}_3\in(\mathbb{F}_{9})^9$, and $k \leq 5$. The retrieval code is chosen as $D_3 = \textnormal{GRS}_4(\boldsymbol{1},\boldsymbol{\alpha}_3)$. We then get $\suco{D_3}{3}$ as a $[9,3,6]$-code with generator matrix
\[
\suco{G_3}{3}=\begin{bmatrix}
1& 0& 0& 2& 1& 2& 2& 0& 1\\
0& 1& 0& 0& 2& 1& 2& 2& 1\\
0& 0& 1& 2& 1& 1& 0& 2& 2 \end{bmatrix}.\]
Its dual is a $[9,6,3]$ code and hence we have a collusion protection of $t'=2$, down from $t=4$ for the scheme $\mathscr{S}$. Calculating the star product codes shows that both schemes have identical rates $R_3 = (6-k)/9$.
\end{exa}

\section{Conclusion \& Future Work}
We have presented an alteration of the PIR scheme of \cite{Gnilke} using subfield subcodes as the retrieval code of the PIR scheme. This scheme achieves a considerable improvement in computational complexity of the server side response calculations while maintaining the same rate. As the exampled show this gain is usually accompanied by a reduction in the collusion resistance.

Future work will explore further examples. For the examples presented it holds that $C \star D = C \star \suco{D}{q}$, hence the rates are equal. Trivial examples for which this equality does not hold are given by codes $D$ of dimension $>1$ for which $\suco{D}{q}$ is the repetition code. These lead to schemes $\suco{\Sch}{q}$ with an increased rate, but at the cost of a complete loss of collusion protection. We will explore the possibility of an example for which the rate increases without the collusion protection completely disappearing. Furthermore, a non-trivial example of a scheme using corollary \ref{tracescheme} would be of interest.

Apart from more examples, the possibility of a class of GRS codes for which the subfield subcodes and their parameters are predictable should be considered. In \cite{Hernando} the authors present some classes of GRS codes with high dimensional subfield subcodes and algorithms to search them.


\begin{thebibliography}{1}

\bibitem{Julien}
Lavauzelle, Julien. \emph{Private information retrieval from transversal designs} IEEE Transactions on Information Theory 65.2 (2018): 1189-1205

\bibitem{Eitan}
  Y. Zhang, E. Yaakobi, T. Etzion, M. Schwartz, \emph{On the Access Complexity of PIR Schemes}, 2019 IEEE International Symposium on Information Theory (ISIT),
  pp.2134-2138
  
  \bibitem{Gnilke}
R. Freij-Hollanti, O. W. Gnilke, C. Hollanti, and D. A. Karpuk. \emph{Private Information Retrieval From Coded Databases With Colluding Servers}, In: SIAM Journal of Applied Algebra and Geometry, vol. 1, pp. 647-664, 2017.

\bibitem{Delsarte}
P. Delsarte. \emph{On Subfield Subcodes of Modified Reed-Solomon Codes}, In: IEEE Transactions on Information Theory, vol. IT-21, pp. 575-576, 1975.

\bibitem{vsGathen}
J. von sur Gathen and J. Gerhard. \emph{Modern Computer Algebra}, 3rd edition, Cambridge University Press, 2014.

\bibitem{Hernando}
F. Hernando, K. Marshall, and M. E. O'Sullivan. \emph{The Dimension of Subcode-Subfields of Shortened Generalized Reed-Solomon Codes}, In: Des. Codes Cryptogr. (2013) 69:131–142.



\end{thebibliography}
\end{document}